\documentclass[prb, twocolumn, showpacs, amsmath, amssymb,superscriptaddress]{revtex4-1}

\usepackage{amssymb,amsfonts,amsmath} 
\usepackage{graphicx}
\usepackage{placeins}
\usepackage{hyperref}
\newcommand{\me}{\mathrm{e}}
\newcommand{\mi}{\mathrm{i}}

\renewcommand{\Re}{\mathrm{Re}}
\newcommand{\diffd}{\mathrm{d}}
\newcommand{\Chi}{\mathrm{X}}
\usepackage{bbm}

\bibpunct{[}{]}{,}{n}{}{}

\begin{document} 

\title{Detecting phases in one-dimensional many-fermion systems with the functional renormalization group}

\author{L.\ Markhof}  
\affiliation{Institut f{\"u}r Theorie der Statistischen Physik, RWTH Aachen University 
and JARA---Fundamentals of Future Information
Technology, 52056 Aachen, Germany}

\author{B.\ Sbierski}  
\affiliation{Dahlem Center for Complex Quantum Systems and Institut f{\"u}r Theoretische Physik,
Freie Universit{\"a}t Berlin, 14195, Berlin, Germany}

\author{V.\ Meden} 
\affiliation{Institut f{\"u}r Theorie der Statistischen Physik, RWTH Aachen University 
and JARA---Fundamentals of Future Information
Technology, 52056 Aachen, Germany}

\author{C.\ Karrasch}  
\affiliation{Dahlem Center for Complex Quantum Systems and Institut f{\"u}r Theoretische Physik,
Freie Universit{\"a}t Berlin, 14195, Berlin, Germany}

\begin{abstract} 

The functional renormalization group (FRG) has been used widely to investigate phase diagrams, in particular the one of the two-dimensional Hubbard model. So far, the study of one-dimensional models has not attracted as much attention. We use the FRG to investigate the phases of a one-dimensional spinless tight-binding chain with nearest and next-nearest neighbor interactions at half filling. The phase diagram of this model has already been established with other methods, and phase transitions from a metallic phase to ordered phases take place at intermediate to strong interactions. The model is thus well suited to analyze the potential and the limitations of the FRG in this regime of interactions. We employ flow equations that are exact up to second order in the interaction, which implies that we take into account the frequency dependence of the two-particle vertex as well as the feedback of the dynamic self-energy. For intermediate nearest neighbor interactions, our scheme captures the phase transition from a metallic phase to a charge density wave with alternating occupation. The critical interaction, at which this transition occurs, is underestimated due to our approximations. Similarly, for intermediate next-nearest neighbor interactions, we observe a transition to a charge density wave with occupation pattern $..00110011..$. We show that taking into account a feedback of the two-particle vertex in the flow equation is essential for the detection of those phases.

\end{abstract}

\pacs{} 
\date{\today} 
\maketitle

\section{Introduction}
\label{sec:introduction}

The functional renormalization group (FRG) is a versatile tool to treat many-body systems with diverse energy scales and competing ordering tendencies \cite{metzner2012,kopietz2010,andergassen2008}. In this particular flavor of the RG concept, a flow parameter $\Lambda$ is introduced in such a fashion that at an initial $\Lambda_\text{i}$ the system can be solved exactly. By successively eliminating this cutoff one then recovers the full interacting problem, summing up all Feynman diagrams for, e.g., the one-particle irreducible interacting vertex functions. In practice, the infinite hierarchy of flow equations emerging from the formalism has to be truncated. The resulting coupled differential equations can then be solved, usually numerically, to obtain approximations for, e.g., the self-energy. 

The FRG can be used for many different applications \cite{metzner2012,kopietz2010}, but in this paper we focus on the study of phase transitions. In fact, the FRG for quantum many-body systems was introduced specifically for this purpose several years ago. It was used to examine the phase diagram of the two-dimensional  Hubbard model at weak coupling \cite{halboth2000}. Numerous authors have extended and refined the treatment since then, for example in  Refs.~\cite{honerkamp2001, salmhofer2004, baier2004, gersch2005, husemann2012, giering2012, uebelacker2012, eberlein2014, eberlein2015, seiler2016, volpez2016, delapena2017, vilardi2017}. Depending on the geometry, the interaction strength, and doping, different leading-order instabilities were identified.

In one dimension, the FRG has not been employed as extensively to study phases; in Ref.~\cite{tam2006}, the half-filled extended Hubbard model at small interactions has been examined. 

As mentioned above, only a few of the infinite number of the FRG flow equations can actually be taken into account in a numerical computation. Neglecting the flow of the vertex functions of order $m+1$ and higher leads to a scheme where all Feynman diagrams of order $m$ in the interaction are included. To give an example, neglecting the flow of the two-particle vertex gives a result for the self-energy that is at least as good as first order perturbation theory. The standard truncation thus renders FRG a weak coupling scheme. However, the FRG often performs well even at intermediate interactions since an infinite number of diagrams is resummed during the flow \cite{metzner2012,kopietz2010}. Furthermore, due to the scale-dependent treatment the FRG manages to cure infrared divergences that afflict the perturbative approach. Besides, no bias is introduced when treating competing instabilities in contrast to, e.g., mean-field approaches. Next to those $m$th order truncation schemes, other approximations are possible. In most of the studies of phase diagrams in both one and two dimensions, only the flow of the (static) two-particle vertex was considered, neglecting the self-energy feedback \cite{metzner2012}. The effect of discarding this feedback is still under debate today, see Sec.~\ref{subsec:FRG_gen}. Note that those approximations do not follow the inherent construction of truncating the flow equations at a certain order anymore.

In this paper, we use an FRG scheme that follows the intrinsic truncation structure and takes into account all terms up to second order in the interaction \cite{weidinger2017}. It can be formulated in real space as well as in momentum space \cite{sbierski2017}. With this scheme, we study the phase diagram of the one-dimensional spinless tight-binding model with nearest and next-nearest neighbor interactions. The phases of this model are well understood from previous works (see Sec.~\ref{sec:model}), with phase transitions occurring at intermediate to strong interactions. It is thus possible to benchmark our FRG results and to determine whether we are able to capture the different phases. We find that the truncated FRG can provide a qualitative understanding of charge-ordered phases occurring at intermediate interactions, but the quantitative predictions are not reliable. The bond order phase for an intermediate next-nearest neighbor interaction strength (see Sec.~\ref{sec:model}) cannot be identified unambigously from our calculations, which shows that the truncated FRG can miss ordering tendencies.

This paper is organized as follows: In Sec.~\ref{sec:model} we describe our model and summarize previous findings. In Sec.~\ref{sec:FRG} we briefly recapitulate earlier works about phase diagrams using the FRG and outline the key points of the FRG approach used in this paper. A more detailed derivation of the flow equations employed here can be found in the \hyperref[app:detail_FRG]{Appendix}. We present the results of our calculations in Sec.~\ref{sec:results}, where we study the various phases in the parameter regime of repulsive interactions. We conclude in Sec.~\ref{sec:conclusion}.

\section{Model}
\label{sec:model}

\begin{figure}
	\includegraphics[scale=1]{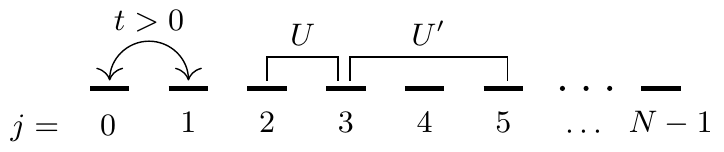}
	\caption{Sketch of the one-dimensional tight-binding chain with the hopping $t$, the nearest neighbor interaction $U$, and the next-nearest neighbor interaction $U^\prime$.}
	\label{fig:tbmodel_sketch}
\end{figure}

We consider a one-dimensional chain of spinless fermions as sketched in Fig.~\ref{fig:tbmodel_sketch} at $T=0$ and at half filling, i.e., on average one fermion per two sites.
The Hamiltonian is given by
\begin{align}
\label{eq:Hamiltonian}
	H =& - t \sum_j \left( c_j^\dagger c_{j+1} +c_{j+1}^\dagger c_j \right)  \notag\\
	 &+ U \sum_j  \left( n_j-\frac{1}{2}\right) \left( n_{j+1}-\frac{1}{2}\right) \notag \\
	 &+ U^\prime \sum_j  \left( n_j-\frac{1}{2}\right) \left( n_{j+2}-\frac{1}{2}\right).
\end{align}
We study both open boundary conditions (OBC) and periodic ones (PBC). For OBC, the first two sums run from $0$ to $N-2$, and the last one from $0$ to $N-3$. For PBC, we identify $N+j$ with $j$, and all sums run from $0$ to $N-1$. We have written the Hamiltonian in a particle-hole symmetric form, such that the chemical potential is fixed at $\mu =0$. In the following, we will set $t=1$ unless otherwise mentioned and thus measure energies with respect to the bare hopping. 

For $U^\prime = 0$, the model is Bethe ansatz solvable \cite{yang1966,yang1966a}. The exact solution shows that the system is in a metallic Luttinger liquid (LL) state, where no symmetry is broken, for $|U| \leq 2$. For $U>2$ a charge density wave with a unit cell of one occupied and one unoccupied site forms (CDW-I). For $U<-2$ the system is in a phase separated state.

In a mean field approximation of the model with $U^\prime=0$, the system is in an ordered CDW-I state for any $U>0$. For small $U$, the self-consistency equation can be solved analytically, and the result for the order parameter is $8/\pi \, (U+\pi)/U \, \exp\{-(U+\pi)/U\}$. A comparison with the exact solution, in which the CDW-I phase is found only for $U>2$, shows that mean field theory is inappropriate to study the phase diagram of the $U^\prime=0$ model. 

Next, let us consider the model with $U^\prime \neq0$. In the atomic limit ($t=0$), the system has two phases for repulsive interactions. For $U^\prime < U/2$, we obtain a CDW-I, whereas for $U^\prime>U/2$ the system is in a charge-ordered state with a unit cell of two occupied and two unoccupied sites (CDW-II).

\begin{figure}
	\includegraphics[scale=1]{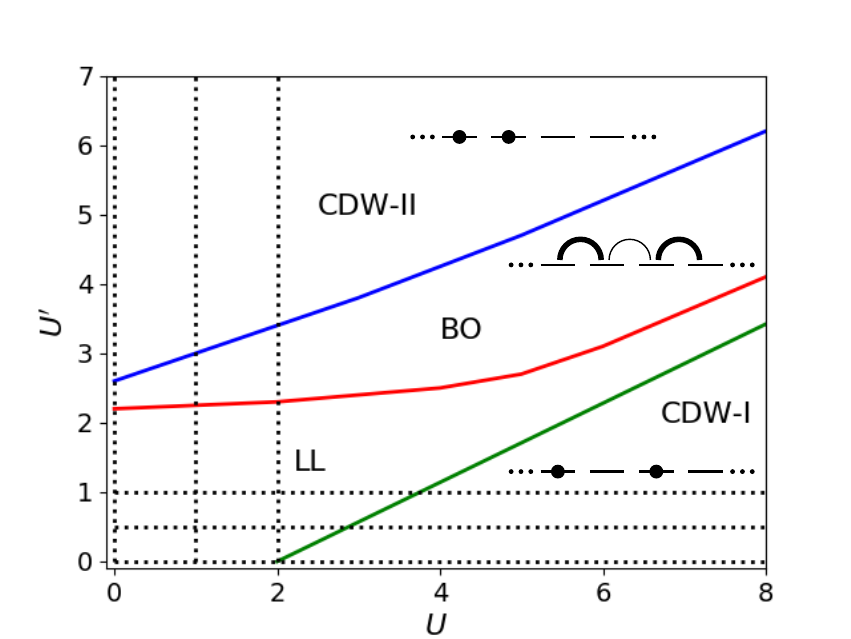}
	\caption{Phase diagram of the model Eq.~\eqref{eq:Hamiltonian} as reported by Mishra \textit{et al}. The above plot was sketched after reading off the approximate location of the transition lines from Fig.~1 of Ref.~\cite{mishra2011}. The full lines indicate the phase transitions, see the text for further details. The sketches indicate the fermion distribution on the sites in the CDW phases or the strength of the bonds in the BO phase. The dashed lines show the cuts of the phase diagram studied in this paper.}
	\label{fig:sketch_phasediag}
\end{figure}

For the full model, no exact solution is known. Its phase diagram has been studied for several decades with various methods, among them a mixture of the renormalization group and the exact diagonalization of small chains \cite{emery1988}, the modified Lanczos method \cite{hallberg1990,zhuravlev1997}, exact diagonalization techniques \cite{poilblanc1997}, and the density matrix renormalization group (DMRG) \cite{schmitteckert2004,mishra2011}. For repulsive interactions, four different phases were found. The phase diagram is sketched in Fig.~\ref{fig:sketch_phasediag}, which qualitatively reproduces Fig.~1 of the paper by Mishra \textit{et al.} \cite{mishra2011}. To prepare Fig.~\ref{fig:sketch_phasediag}, we read off the various critical interaction strengths from Fig.~1 of Ref.~\cite{mishra2011}. If the kinetic energy [first line of Eq.~\eqref{eq:Hamiltonian}] is dominant, the system is in the metallic LL phase. As discussed above, if the nearest neighbor interactions are tuned up a phase transition to the CDW-I takes place. The critical value $U_c$, which is $2$ for $U^\prime=0$, increases with rising $U^\prime$ since the CDW-I becomes less favorable in a system with repulsive next-nearest neighbor interactions. For intermediate $U^\prime$, a bond order phase (BO) emerges due to the competition between the kinetic energy and $U^\prime$. In this phase, the hybridization $\langle c_i^\dagger c_{i+1} \rangle$ oscillates from bond to bond. Finally, when the next-nearest neighbor interaction is increased further the system crosses over to a CDW-II phase. While the CDW phases are physically very intuitive, the BO phase is more difficult to interpret and has been missed in some studies \cite{zhuravlev1997,poilblanc1997}.

In contrast to the two-dimensional Hubbard model \cite{halboth2000} and the extended one-dimensional Hubbard model at half filling \cite{tam2006}, where ordering tendencies are expected as soon as a small interaction is switched on, we here face the challenge of detecting phases in the regime of intermediate couplings. 

\section{Functional Renormalization Group}
\label{sec:FRG}

\subsection{General discussion}
\label{subsec:FRG_gen}
In the FRG a flow parameter $\Lambda$ is introduced, usually in the single-particle Green's function, that allows for a smooth interpolation from an exactly solvable system to the fully interacting model of interest. A convenient choice for the flowing quantities are the one-particle irreducible vertex functions which are the self-energy, the two-particle vertex and so on \cite{metzner2012}. However, following the exact flow would require the solution of an infinite number of coupled differential equations, which is impossible in most cases. Instead, only the flow of a few vertex functions can be taken into account. When interested in the renormalization of both the self-energy and the two-particle vertex, the first approximation is to set the three-particle vertex to its initial value zero, i.e., performing a truncation at the level $m=2$. We then obtain a closed system of coupled differential equations for the self-energy and the two-particle vertex that is exact up to second order in the interaction. 

This system of differential equations is still difficult to solve numerically since the self-energy depends on one frequency and two real space variables (or one momentum in the translationally invariant case), and the two-particle vertex depends on three frequencies and four real space variables (or three momenta). In the early works about the phase diagram of the two-dimensional Hubbard model, additional approximations were implemented after the truncation to keep the computational effort manageable \cite{salmhofer2004}. Only the flow of the static two-particle vertex was considered, discarding the self-energy feedback. The FRG was formulated in momentum space, and due to the specific way of discretizing the Brillouin zone this approach was called the ``$N$-patch FRG'' \cite{honerkamp2001}.  To study the phases of the model, the flow equations were solved until a divergence, termed the flow to strong coupling, was observed at a scale $\Lambda_\star$. Analyzing which component of the vertex function or susceptibility diverged first led to a prediction of the nature of the ordered phase as well as a ``phase diagram''. The same approximation was employed to study the extended Hubbard model in one dimension \cite{tam2006}. 

In later studies of the two-dimensional Hubbard model, the above described method was extended to access also the scales below $\Lambda_\star$ \cite{salmhofer2004,baier2004,gersch2005,eberlein2014}. For a review of the results until 2012, see Ref.~\cite{metzner2012}. The phase diagram of this model is still an active field of research, see, e.g., Refs.~\cite{husemann2012, giering2012, uebelacker2012,eberlein2015,seiler2016,volpez2016,delapena2017,vilardi2017} for various refined and extended treatments. Due to the complexity of the flow equations, so far it has not been possible to implement an FRG scheme that takes the full frequency dependence, the full momentum dependence and the self-energy flow into account.  Although a mostly unified picture concerning the leading order instabilities in the various parameter regimes of the two-dimensional Hubbard model has emerged over the years, it is still not completely clear how the different approximations affect the results. 

In this paper, we apply a scheme which follows the logic of an $m=2$ order truncation. The overall concept of this approach has first been reported in Ref.~\cite{bauer2014}, building up on ideas of Refs.~\cite{karrasch2008,jakobs2010}, and has been extended in Ref.~\cite{weidinger2017}. It is applicable for Hamiltonians with arbitrary single-particle terms and finite-ranged interactions. The authors of Ref.~\cite{bauer2014} called this flavor of the FRG ``coupled ladder approximation'' and in Ref.~\cite{weidinger2017} ``extended coupled ladder approximation'' (eCLA). They were mostly interested in the transport through quantum point contacts, calculating the conductance and the susceptibility, with special focus on the so-called ``0.7-anomaly'' \cite{bauer2013}. As the paper by Weidinger \textit{et al.} \cite{weidinger2017} contains a detailed derivation of the flow equations in the eCLA for a one-dimensional model including spin, in this section we only mention the main steps and introduce the flowing quantities for our spinless model. More details can be found in the \hyperref[app:detail_FRG]{appendix}.

The Hamiltonian Eq.~\eqref{eq:Hamiltonian} is formulated in real space. For a translationally invariant system, it can be advantageous to go to momentum space. Also in this case, second order FRG flow equations can be derived, which is covered in detail in Ref.~\cite{sbierski2017}. We therefore do not write down the momentum space equations here. Reference \cite{sbierski2017} also contains real space flow equations which are similar, but not equal to the ones we use here, see below. In Ref.~\cite{sbierski2017}, the authors used the real space approach to study the transport with abrupt junctions, and the momentum space approach to examine LL power-law behavior in the occupation as a function of the momentum. Here, we employ the momentum space FRG in addition to the real space scheme and compare the results for phase transitions in Sec.~\ref{subsec:cdwi}.

\subsection{FRG flow equations}
\label{subsec:FRG_floweqs}
\begin{figure}
	\includegraphics[scale=1,trim={0 0 5cm 0},clip]{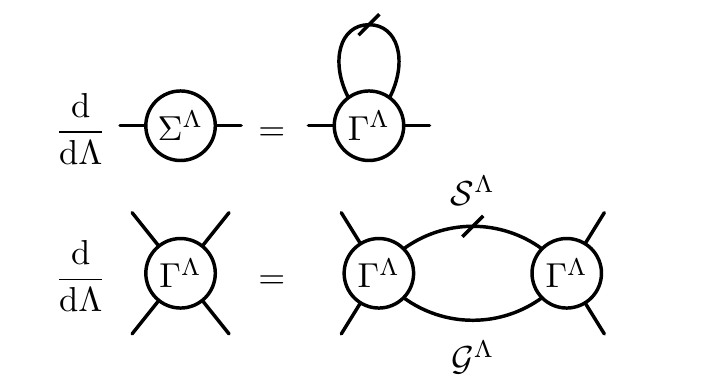}
	\caption{Flow equations for the self-energy $\Sigma^\Lambda$ and the two-particle vertex $\Gamma^\Lambda$, neglecting the three-particle vertex.}
	\label{fig:floweq_diag}
\end{figure}

Let us now recapitulate the main steps in the derivation of the eCLA. The flow equations for the self-energy and the two-particle vertex in a diagrammatic language after setting the three-particle vertex to zero are depicted in Fig.~\ref{fig:floweq_diag}.
Following the logic of the FRG, the full two-particle vertex is present on the right-hand side of the flow equation. We now note that the two-particle vertex can be naturally divided in different parts \cite{karrasch2008}, compare the structure of the general flow equation for the two-particle vertex Eqs.~\eqref{eq:flowGammap_general} - \eqref{eq:flowGammad_general}. Then,
\begin{equation}
	\Gamma^\Lambda(j_1^\prime, j_2^\prime; j_1, j_2; \Pi, \Chi, \Delta) = V + \Gamma^\Lambda_p + \Gamma^\Lambda_x + \Gamma^\Lambda_d,
\end{equation}
with the bare interaction $V$, the particle-particle channel $\Gamma^\Lambda_p$, the exchange (sometimes also called crossed) particle-hole channel $\Gamma^\Lambda_x$ and the direct particle-hole channel $\Gamma^\Lambda_d$. For our model Hamiltonian Eq.~\eqref{eq:Hamiltonian}, $V$ has as non zero components
\begin{align}
	&V(j,j+ 1; j, j + 1)  = U, \\
	&V(j,j+ 2; j, j+ 2) = U^\prime,
\end{align}
and the corresponding ones from permutation of the indices ($\Gamma$ is antisymmetric under the exchange of the $j_1^\prime, \ j_2^\prime$ or the $j_1,\ j_2$ index pair).

The key point in the eCLA is to insert (at first) only the bare vertex $V$ instead of the full $\Gamma^\Lambda$ on the right-hand side of the flow equation. The neglected terms are all at least of third order in the interaction. In a perturbative sense, our approximation is thus as good as it was before, when we neglected the flow of the three-particle vertex. It is then possible to restrict the variable dependence of the vertex channels such that each depends on only one bosonic frequency. In addition, due to the finite range of the bare interaction, the dependence on four real-space indices can be limited as well. In compact notation, we can now write the vertex channels as
\begin{align}
	P_{i,j}^{k,l;\Lambda}(\Pi)&  := \Gamma^\Lambda_p(i,i+k; j, j+l; \Pi) \label{eq:defP}\\
	X_{i,j}^{k,l;\Lambda}(\Chi) & := \Gamma^\Lambda_x(i,j+l;j,i+k; \Chi) \label{eq:defX}\\
	D_{i,j}^{k,l;\Lambda}(\Delta) & := \Gamma^\Lambda_d(i,j+l; i+k,j; \Delta). \label{eq:defD}
\end{align}
The subscript indices $i$ and $j$ range in principle between $0$ and $N-1$ (see the \hyperref[app:detail_FRG]{appendix} for further details). The superscript indices lie in $[-L_U, L_U]$, where $L_U$ is the spatial range of the bare interaction. For the Hamiltonian Eq.~\eqref{eq:Hamiltonian}, $L_U=2$ (or $L_U=1$ if $U^\prime=0$). 

The above described approximation scheme leads to the flow equations that were used in Ref.~\cite{sbierski2017} (note that the notation in Ref.~\cite{sbierski2017} is slightly different from the one used here).

As a next step, it is also possible to include a vertex feedback without destroying the above structure of the channels \cite{jakobs2010}, which along the flow leads to components in the vertices with $|k|, |l| > L_U$. Let us first consider the particle-particle channel. Instead of inserting the bare interaction on the right-hand side of the flow equation for $\Gamma_p^\Lambda$, we replace
\begin{align}
\label{eq:defGammaptilde}
		\Gamma^\Lambda (j_1^\prime, j_2^\prime; j_1, j_2; \Pi, \Chi, \Delta) \rightarrow  \tilde{\Gamma}_p^\Lambda (j_1^\prime, j_2^\prime; j_1, j_2; \Pi) \notag \\
	  :=  \delta_{j_1^\prime, j_2^\prime}^L \delta_{j_1,j_2}^L \Gamma^\Lambda (j_1^\prime, j_2^\prime; j_1, j_2; \Pi,0,0),
\end{align}
where $\delta_{i,j}^L = 1 $ if $|i-j| \leq L$ and zero otherwise. $\tilde{\Gamma}_p$ contains the bare interaction as well as all components of $P^\Lambda(\Pi)$. Besides, also components of $X^\Lambda(0)$ and $D^\Lambda(0)$ are fed back in the flow equation. The latter channels are taken at zero frequency to avoid a mixing of the three bosonic frequencies. We have now introduced the \textit{feedback length} $L$, which restricts the upper indices in $P_{i,j}^{k,l}$ to lie in the range $[-L,L]$. 

With an analogous consideration for the other channels, we arrive at a vertex that instead of $\mathcal{O}(N_f^3 \, N^4)$, with $N_f$ the number of discretized frequencies (see Sec.~\ref{subsec:num_impl}), has only $\mathcal{O}(N_f \, N^2 L^2)$ variables. If we wanted to compute the full real-space dependence, we would have to take $L\approx N/2$ for PBC and $L=N-1$ for OBC. This parametrization is thus only advantageous if $L$ can be chosen much smaller. Fortunately, this is indeed the case for our model as we will see in Sec.~\ref{sec:results}.

 At each scale, we insert the dynamic vertex  on the right-hand side of the flow equation for the self-energy, which becomes dynamic as well. We stress that with this strategy, our scheme is exact up to second order in the interaction. The flow equation for the self-energy now reads
\begin{align}
	 & \frac{\diffd}{\diffd \Lambda} \Sigma^\Lambda_{i,j} (\omega)  \notag \\
	  &= -\frac{1}{\beta} \sum_{i^\prime, j^\prime, \omega^\prime} \mathcal{S}^\Lambda_{j^\prime, i^\prime}(\omega^\prime) \Big[  V(i,i^\prime; j,j^\prime)  + \Gamma^\Lambda_p(i,i^\prime; j,j^\prime; \omega + \omega^\prime) \notag \\
	   & \hspace{4em} + \Gamma^\Lambda_x(i,i^\prime; j,j^\prime;\omega-\omega^\prime)   + \Gamma^\Lambda_d(i,i^\prime; j,j^\prime;0) \Big],
\end{align} 
where $\mathcal{S}^\Lambda$ is the single-scale propagator, defined as the derivative of the full propagator with respect to the flow parameter $\Lambda$ at fixed self-energy.  The initial condition for the sharp cutoff specified in Eq.~\eqref{eq:cutoff} is for finite $\Lambda_i$ given by $\Sigma^{\Lambda_\text{i}} = 0$ (see the \hyperref[app:detail_FRG]{appendix}). However, later we will also add small symmetry breaking terms in the initial condition to induce a phase transition.

The flow equations for the vertex channels can be written in a compact matrix notation (we do not write the flow equation for $D$ since due to symmetry $D=-X$) as 
\begin{align}
	\frac{\diffd}{\diffd \Lambda} P^\Lambda (\omega) & = \tilde{P}^\Lambda(\omega) \cdot W^{p \Lambda}(\omega) \cdot \tilde{P}^\Lambda(\omega) \label{eq:flowP}\\
	\frac{\diffd}{\diffd \Lambda} X^\Lambda (\omega) & = \tilde{X}^\Lambda(\omega) \cdot W^{x \Lambda}(\omega) \cdot \tilde{X}^\Lambda(\omega). \label{eq:flowX}
\end{align}
The dot represents a block-matrix multiplication in each index, $[A\cdot B]_{i,j}^{k,l} = \sum_{i^\prime  k^\prime } A_{i, i^\prime}^{k,k^\prime} B_{i^\prime, j}^{k^\prime,l}$. The tilde above the vertices indicates that the bare vertex as well as dynamic feedback from the same channel and static feedback from the other channels is included, see Eqs.~\eqref{eq:defP} and \eqref{eq:defGammaptilde} for the definition of $\tilde{P}$. The full expressions for $\tilde{P}$ and $\tilde{X}$ for our model can be found in Eqs.~\eqref{eq:Ptilde} and \eqref{eq:Xtilde} in the appendix. The matrices $W^{x/p \Lambda}$ represent a product of a single-scale propagator and a full propagator, summed over an internal frequency. The initial conditions are given by $P^{\Lambda_\text{i}} = X^{\Lambda_\text{i}} = 0$.

We introduce the FRG flow parameter by multiplying a sharp cutoff function to the free Green's function,
\begin{equation}
\label{eq:cutoff}
	\mathcal{G}^\Lambda_0 (\omega) = \Theta(|\omega|-\Lambda) \mathcal{G}_0 (\omega).
\end{equation}
Then, a $\delta$-distribution $\delta(| \omega| -\Lambda)$ appears in the single-scale propagator, which makes it possible to analytically evaluate the frequency integrals. The resulting expressions for the matrices $W^{x/p \Lambda}$ are given in Eqs.~\eqref{eq:def_Wp}-\eqref{eq:def_Wx0}, and the final flow equation for the self-energy in Eqs.~\eqref{eq:flowS} and \eqref{eq:flowS0}.

The flow equations described above are still complicated. They can be simplified by applying the static approximation \cite{bauer2014}, in which only the frequency $\omega=0$ is considered.  Note that this static approximation does no longer contain all second order diagrams for the static self-energy. Thus, also observables calculated from $\Sigma^{\Lambda_\text{f}}$ are only exact up to first order.

\subsection{Numerical implementation}
\label{subsec:num_impl}

The flow parameter $\Lambda$ we introduced in the free Green's function ranges from infinity to zero. However, for computational purposes we have mapped this semi-infinite interval to $(1,0]$ by going to $x = \Lambda/(1+\Lambda)$ \cite{bauer2014}. We usually start at $x_\text{i}=0.999999$, which corresponds to $\Lambda_\text{i} \approx 10^6$, and have tested that starting the integration at $x_\text{i}$ closer to one does not change the results. For the integration of the system of coupled differential equations we use a standard Runge-Kutta algorithm with adaptive step-size control.

For the dynamic calculation, we have to discretize the continuous frequency $\omega$. We will always use a geometrically spaced grid with
\begin{equation}
	\omega_n = \omega_1 \frac{a^n-1}{a-1}, \quad n = 0 .. N_f-1, \ \omega_1 > 0, \ a > 1.
\end{equation}
If a quantity is needed at a frequency in between the discrete grid points we use linear interpolation. A resolution with $N_f = 60, \ \omega_1 = 0.001$, and $a=1.2$ is in our case sufficient to obtain converged results. We also tested that calculations with a logarithmically spaced frequency grid do not change the outcome.

Since we consider the half-filled case, we will always use an even number of sites to ensure the correct filling. In addition, we have to consider systems of size $N=4 k+2$ in the case of PBC. This is because the free system has eigenenergies $e_n = -2 \cos(k_n), \ k_n = 2 \pi n / N, \ n \in (-N/2, N/2] $. If $N$ is a multiple of four, one eigenvalue is zero, and the free Hamiltonian can not be inverted.

The computationally most expensive parts about the above presented algorithm are the matrix-matrix multiplications in Eqs.~\eqref{eq:flowP} and \eqref{eq:flowX}, since they scale as $\mathcal{O}(N^3 L^3)$. In a dynamic scheme, Eqs.~\eqref{eq:flowP} and \eqref{eq:flowX} have to be computed for each of the $N_f$ frequencies. We also have to invert matrices of size $N \times N$ to get the propagator Eq.~\eqref{eq:propagator} at various frequencies. We were able to get results for systems up to length $\sim 400$ in the static approximation using highly optimized BLAS routines with shared-memory parallelization. For the dynamic case, MPI parallelization makes it possible to go up to system sizes of $\sim 80$. Due to the structure of the flow equations, it is comparatively easy to parallelize the code using MPI. In our code, each MPI rank calculates the right-hand side for a share of frequencies (this limits the number of nodes that can be used to $N_f$), which leads to an even workload distribution.

\subsection{Observables}
After the integration of the flow equations down to $\Lambda_\text{f}=0$, we obtain an approximation for the interacting self-energy and vertex at discrete frequencies. We will later be interested in the occupation and the hybridization, so we briefly explain how to obtain expectation values of $c_i^\dagger c_j$ from $\Sigma^{\Lambda_\text{f}}$, or rather from the full propagator [see Eq.~\eqref{eq:propagator}] at $\Lambda_\text{f}$. For a dynamic self-energy, we have to solve a frequency integral 
\begin{equation}
	\langle c_i^\dagger c_j \rangle = \frac{1}{2 \pi} \left[ \int_{-\infty}^\infty \me^{\mi \omega 0^+} \, \tilde{\mathcal{G}}^{\Lambda_\text{f}}(\omega) \right]_{i,j}.
\end{equation} 
We use a quadrature integration routine, and approximate the self-energy in between the known discrete frequencies  with linear interpolation. The factor $\me^{\mi \omega 0^+}$ is only relevant at very large $\omega$, and can be discarded when adding $\frac{1}{2} \delta_{ij}$ to the result of the integral in numerical computations. For the static approximation, we consider the self-energy as an effective single-particle potential. Then, diagonalizing the kinetic part of the Hamiltonian, $H_0$, together with $\Sigma^{\Lambda_\text{f}}$, we can calculate expectation values from
\begin{equation}
	\langle c_i^\dagger c_j \rangle = \sum_{k} \Theta(-\lambda_k) O_{ik} O_{jk}
\end{equation}
with $\lambda_k$ the eigenvalues of $\tilde{H} = H_0 + \Sigma^{\Lambda_\text{f}}$ and $O$ the orthogonal matrix that diagonalizes $\tilde{H}$.

\section{Results}
\label{sec:results}

In this section, we present results of our second order FRG calculations and give a detailed description of how we detect a phase transition. Unless otherwise stated, we always use the real space scheme described in Sec.~\ref{subsec:FRG_floweqs}. In Sec.~\ref{subsec:cdwi}, we also compare to the momentum space approach.

\subsection{Comparison to exact diagonalization}
\begin{figure}
	\includegraphics[width=\linewidth]{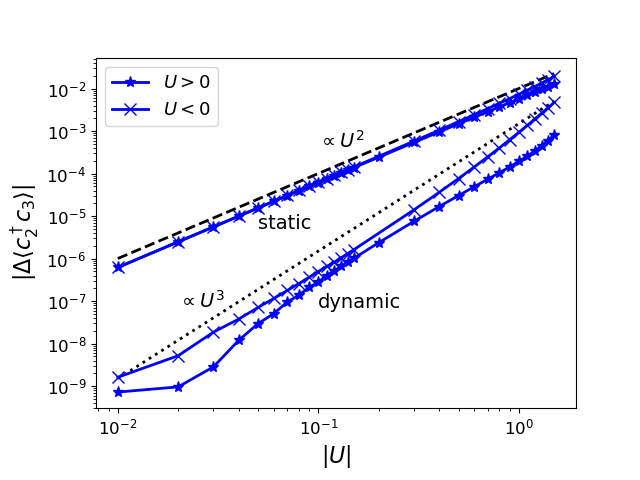}
	\caption{Difference between the ED and the FRG result for the expectation value $\langle c_2^\dagger c_3 \rangle$ for a system with $N=10, \ L= 2, \ U^\prime=0$.}
	\label{fig:FRG_relerr}
\end{figure}

A good check for the implementation is a comparison with exact diagonalization (ED). For very small systems, it is possible to compute the full many-body Hamiltonian in matrix form using all Fock states with $N/2$ occupied and $N/2$ empty sites. We can then diagonalize this large matrix to find the ground state, and calculate expectation values of operators from this. In Fig.~\ref{fig:FRG_relerr} the absolute difference between the ED and the FRG is shown for the hybridization $\langle c_2^\dagger c_3 \rangle$ for a system with PBC, $N=10$ sites and no next-nearest neighbor interaction $U^\prime=0$. We plot $| \langle c_2^\dagger c_3 \rangle_\text{ED} - \langle c_2^\dagger c_3 \rangle_\text{FRG} |$ versus the absolute value of the interaction strength. Both repulsive ($U>0$) and attractive ($U<0$) interactions were considered. The dashed line is proportional to $U^2$, and the dotted line is proportional to $U^3$. The difference between the static FRG result and ED is consistent with an error $\propto U^2$ as expected. The full dynamic scheme is consistent with an error $\propto U^3$. The discrepancy between the slope of the dotted line and the curves for very small interactions with $|\Delta \langle c_2^\dagger c_3 \rangle| < 10^{-(7..8)}$ is due to numerical inaccuracies of the FRG results. For larger interactions ($|U| \gtrsim 0.8$), higher order corrections start to matter. Still, the FRG performs well even in this regime, and later on we go up to interactions $U, \ U^\prime \approx3$. 

\subsection{Phase transition to CDW-I}
\label{subsec:cdwi}
We examine the phase transition from the metallic LL phase to the charge density wave with pattern $..1010..$ for PBC. We consider here a fixed $U^\prime$, i.e., we move along a horizontal line in Fig.~\ref{fig:sketch_phasediag}. Let us start with the model with $U^\prime=0$, for which the transition is at $U_c=2$. When searching for phase transitions, commonly the parameter driving the transition is increased until a divergence in a vertex function is detected \cite{metzner2012}. We first use a different approach, but we will return to the usual ansatz below. Here, we add a small initial symmetry-breaking perturbation \cite{salmhofer2004,gersch2005} that nudges the system to a CDW-I, thus choosing in the ordered phase one of the two degenerate ground states. Namely, we impose as an initial condition on the diagonal of the self-energy 
\begin{equation}
	\Sigma_{j,j}^{\Lambda_\text{i}} (\omega_n) = (-1)^j \, S
\end{equation}	
with a small $S$. As we will see below, when doing so we can integrate the FRG flow equations down to $\Lambda_\text{f}=0$ for all $U$. We are interested in the behavior of the CDW-I order parameter
\begin{equation}
	\langle O_\text{cdwi} \rangle = \frac{2}{N} \sum_{j=0}^{N-1} (-1)^{j+1} \langle n_j \rangle.
\end{equation}
We note that this way of detecting a phase transition is only possible if the flow of the self-energy is taken into account.

We will now investigate whether or not the FRG captures the phase transition and in particular address the following questions: (i) How important is it to keep the frequency dependence of the vertex; (ii) can we access the thermodynamic limit -- i.e., can we find convergence in $N$ -- and what is the role of the feedback length $L$; (iii) how does the choice of $S$ affect the results;  (iv) can we draw an unbiased conclusion despite our small initial symmetry breaking term; (v) which feedback terms are necessary in the FRG flow equations to capture the phase transition; (vi) how does the real space scheme compare to the momentum space approach.

\begin{figure}
	\includegraphics[width=\linewidth]{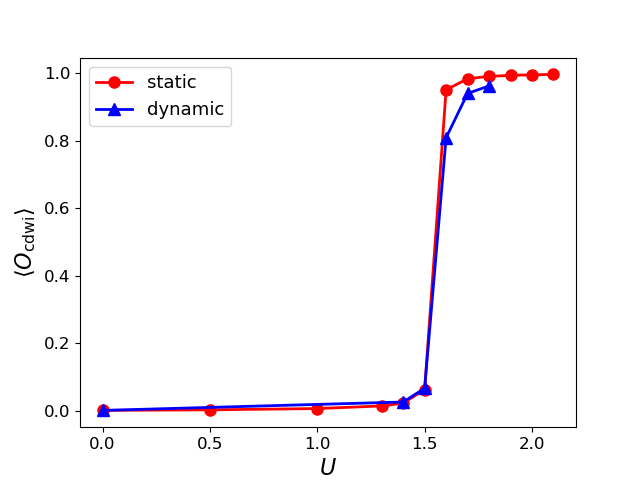}
	\caption{Phase transition to the CDW-I, comparison between the static and the dynamic result for $U^\prime=0, \ N=82,\ L=10, \ S = 0.001$.}
	\label{fig:cdwi_static_dynamic}
\end{figure}

(i) A comparison of the dynamic and the static (frequency dependence neglected completely) calculation is shown in Fig.~\ref{fig:cdwi_static_dynamic}.  As can be seen, the CDW-I order parameter as a function of the interaction strength shows a strong increase from a very small value to nearly one. For $N=82$, the critical interaction at which $\langle O_\mathrm{cdwi} \rangle \gg S $ is given by $U_c = 1.6$.
The results for the static and the dynamic calculation are very similar, and they yield the same $U_c$. Apparently the static approximation is already sufficient to capture the phase transition. Since the dynamic calculation is computationally much more expensive, we will only consider the static scheme for the rest of this subsection.

\begin{figure}
	\includegraphics[width=\linewidth]{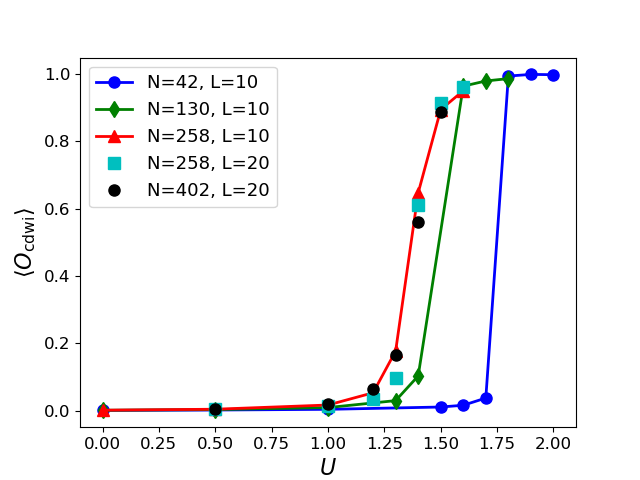}
	\caption{Phase transition to the CDW-I for different system sizes, $U^\prime=0, \ S = 0.001$.}
	\label{fig:cdwi_cmpNL}
\end{figure}

(ii) We now focus on the question of convergence in $N$ and $L$. When comparing small chains, we find that the critical interaction depends on the system length. This is due to finite-size effects and we have to increase $N$ until $U_c$ does not change anymore. Figure~\ref{fig:cdwi_cmpNL} shows that convergence is reached for a system size of $N=258$. We observe that the transition is less sharp for larger system sizes.
Figure~\ref{fig:cdwi_cmpNL} also shows the convergence in the feedback length $L$. The way we set up the real space parametrization of the vertex is only useful if $L$ can be chosen much smaller than $N/2$ (PBC). As can be seen, for our model this is indeed the case. A feedback length of $L=10$ for a system with $N=258$ sites seems to be sufficient to obtain the converged result $U_c = 1.4$. Thus FRG underestimates the critical interaction, since we know that $U_c=2$ from the exact solution. 

\begin{figure}
	\includegraphics[width=\linewidth]{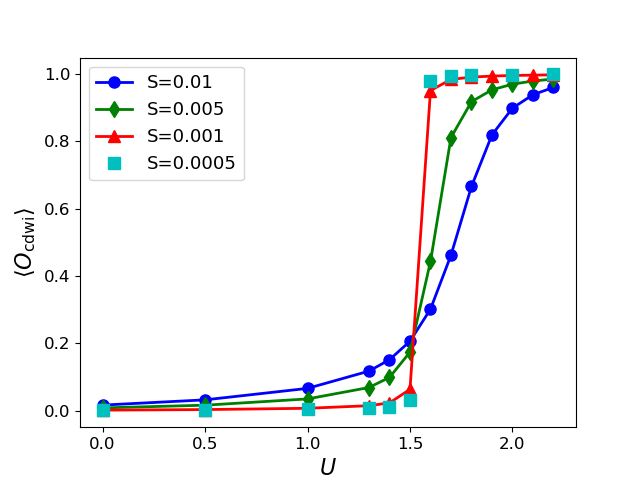}
	\caption{Comparison of different choices of the small initial symmetry breaking term. $U^\prime=0, \ N=82, \ L=10$.}
	\label{fig:cdwi_cmpS}
\end{figure}

(iii) So far, we have only shown results for $S=0.001$. In Fig.~\ref{fig:cdwi_cmpS}, we compare the results for different $S$ for a system with $N=82$ and $L=10$. 
A comparatively large $S$ leads to a smooth, smeared out transition, and for $S=0.01$ and $S=0.005$ the critical interaction cannot be read off easily. At $S=0.001$, the order parameter increases from a small value to nearly one quickly with $U$, and the result for the even smaller $S=0.0005$ confirms this tendency. We know from the exact solution ($S=0$) that the order parameter is exponentially suppressed slightly above $U_c=2$ (BKT transition) \cite{cazalilla2011}. Performing DMRG with an $S$ comparable to the one used here shows that the interaction range over which the order parameter increases from nearly zero to nearly one is much larger than we would conclude from our FRG results.

(iv) Let us now come back to the usual way of detecting a phase transition. Connected to this is the essential question whether the initial condition imposes the final state of the system. If this was the case, our conclusion would be inherently biased. Fortunately, this is not the case here.

\begin{figure}
	\includegraphics[width=\linewidth]{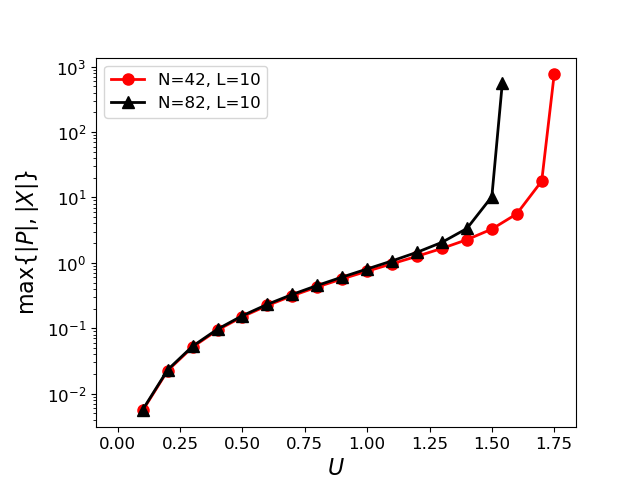}
	\caption{Maximum of the absolute value of all components of $P$ and $X$ without an initial symmetry breaking term. For $U$ larger than shown, it is not possible to solve the flow equations down to $\Lambda_\text{f}=0$.}
	\label{fig:maxPX}
\end{figure}
Figure~\ref{fig:maxPX} shows the maximum of the absolute value of all components of $P$ and $X$ when solving the flow equations without a small initial symmetry breaking term ($S=0$). As can be seen, close to the critical interaction (above which the order parameter became nearly one with a finite $S$), this value becomes very large. For larger $U$, it is not possible to integrate the flow equations down to $\Lambda_\text{f}=0$.

We have also examined what happens if we impose a ``wrong'' initial condition which in this context means that we nudge our system to a BO or a CDW-II state by using appropriate initial conditions. As for the case with $S=0$, the flow equations cannot be solved numerically for $U \geq U_c(N,L)$. An exception occurs when we start with a small symmetry breaking term for a CDW-II with $\Sigma_{j,j}^{\Lambda_\text{i}} = (-1)^{\lfloor (j+1)/2 \rfloor } S$. Since $N/2$ is an odd number (cf. Sec.~\ref{subsec:num_impl}), we favor a distribution with site $0$ unoccupied and site $N-1$ occupied, and in between repeatedly two occupied and two unoccupied sites. This means that, since we consider PBC, an occupied and an unoccupied site are next to each other as favored by the large $U$. In this case, the FRG flows for $U \geq U_c(N,L)$ in the CDW-I phase, ``overriding'' the initial CDW-II bias in between.

The difficulty to obtain results for $\Lambda_\text{f}=0$ for large $U$ without a small correct initial symmetry breaking term could arise for different reasons, namely because of numerical problems, because of our approximations, or because there is some underlying physical instability. From our data, we cannot come to a final conclusion. We cannot rule out that our approximations or numerics (such as problems with an eigenvalue close to zero in the matrix we have to invert ) cause the problem. However, even for $S=0$, the FRG indicates a particular behavior at a certain $U_c(N,L)$. Since imposing the correct small initial symmetry breaking term drives the system into an ordered phase, we believe that the FRG can indeed capture the underlying physical phase transition.

\begin{figure}
	\includegraphics[width=\linewidth]{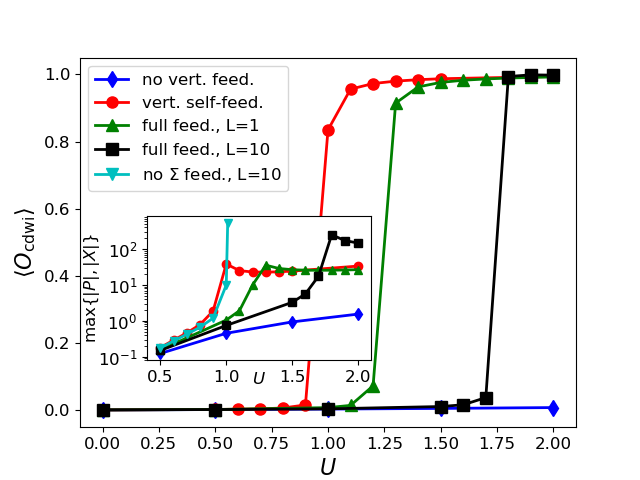}
	\caption{Comparison of the results for the phase transition from the LL phase to the CDW-I for different FRG feedback schemes. $U^\prime=0, \ N=42, \ S= 0.001$. The inset shows the maximum of all components of $P$ and $X$.}
	\label{fig:cdwi_vertfeed}
\end{figure}

(v) The crucial point to detect the phase transition is taking into account the vertex feedback, which can be seen from a comparison to a first-order scheme or a second-order scheme without vertex feedback. In Fig.~\ref{fig:cdwi_vertfeed} we compare the CDW-I order parameter for different vertex feedback schemes for a small chain with $N=42$. Without vertex feedback (i.e., if we only insert the bare vertex on the right-hand side of the flow equations for the vertices) there is no transition to the ordered phase, see the curve labeled as ``no vert. feed.''. Including a feedback of the vertex to itself (i.e., we replace $\Gamma \rightarrow V + \Gamma_x$ on the rhs of the flow equation for vertex $\Gamma_x$, which implies $L=1$, label ``vert. self-feed.'') already leads to the detection of the phase transition, but at a smaller critical interaction. Including also feedback of the other channels (labeled as ``full feedback'', but note that this is still a static calculation) and increasing the feedback length $L$ until we obtain converged results as described above leads us to our final result, $U_c = 1.8$ for $N=42$.

To connect to the studies of the two-dimensional Hubbard model, we have also checked what happens if we neglect the flow of the self-energy. In this case, the expectation values such as $\langle n_i \rangle$ are unchanged from their initial values and an analysis as above is meaningless. Instead, we focus on the maximum of all components of the vertex as in Fig.~\ref{fig:maxPX}. This is shown in the inset of Fig.~\ref{fig:cdwi_vertfeed}. As can be seen, for all schemes that include a self-energy feedback all components of $P$ and $X$ remain finite for all $U$ if we include a small initial bias. This is not the case if we neglect the flow of the self-energy. To obtain the results for the curve labeled as ``no $\Sigma$ feed., $L=10$'', we included the vertex feedback, but the self-energy was not renormalized. Then, for interactions larger as shown, the flow equations could not be integrated down to $\Lambda_\text{f}=0$. Thus also without a self-energy feedback, we find an indication that a phase transition at an intermediate interaction takes place, albeit at a different interaction strength than in our full scheme. However, the components of the two-particle vertex are much harder to interpret compared to the case where we have access to a renormalized self-energy and thus to the order parameter. To gain more insight about the nature of the ordered phase without a flowing self-energy, we would have to implement the flow equations of the susceptibilities as well \cite{halboth2000}. 

\begin{figure}
	\includegraphics[width=\linewidth]{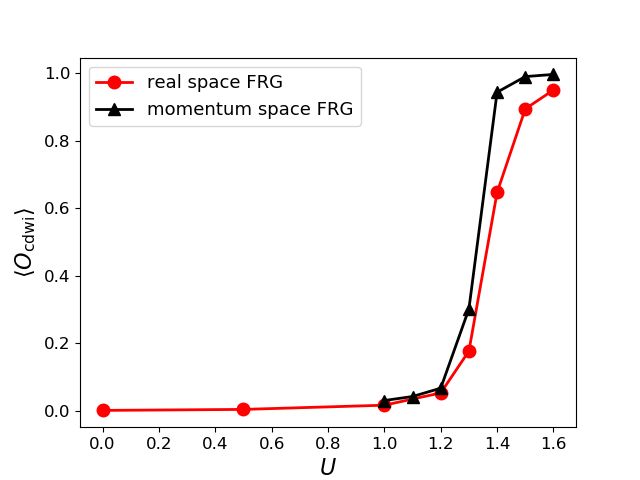}
	\caption{Comparison between the real-space FRG and the momentum space scheme with $S=0.001$. For the real space FRG, the system parameters are $N=258,\ L=10$.}
	\label{fig:cdwi_momentum_realspace}
\end{figure}

(vi) We now turn to a comparison between the real space scheme and the momentum space approach \cite{sbierski2017}. In the latter formulation, the thermodynamic limit can in principle be accessed directly. However, the continuous momentum has to be discretized for a numerical computation, and achieving convergence in the parameters of this momentum grid can be difficult if the discrete $k$-points are not chosen to be equidistant. Figure~\ref{fig:cdwi_momentum_realspace} shows a comparison of the real space and the momentum space results, both in the static approximation. The results agree nicely, in particular the critical interaction $U_c=1.4$ is the same. We attribute the remaining differences around the critical interaction to residual finite-size effects in the real space scheme and numerical inaccuracies.

We conclude that our second order FRG scheme correctly predicts a phase transition to the CDW-I, but the critical interaction strength is underestimated.  In the thermodynamic limit, we obtain $U_c = 1.4$ which is smaller than the true result due to the approximations we have applied. Besides, the order parameter is nearly one already slightly above the critical interaction, instead of increasing very slowly (exponentially suppressed), which would be expected from the exact solution.

\begin{figure}
	\includegraphics[width=\linewidth]{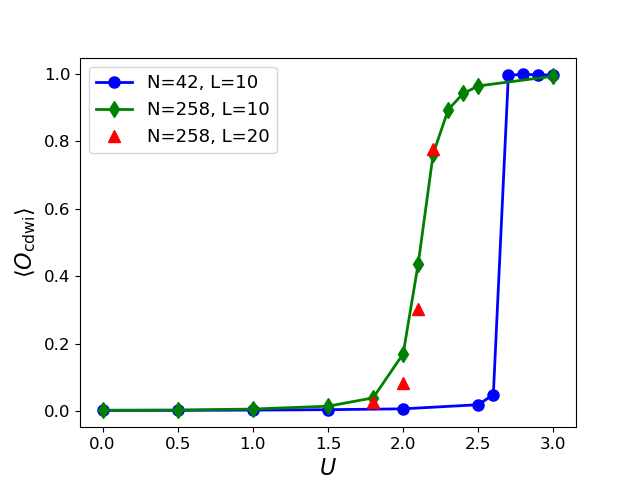}
	\caption{Phase transition LL - CDW-I for $U^\prime=0.5, \ S=0.001$.}
	\label{fig:cdwi_Up0.5}
\end{figure}

We now turn to $U^\prime>0$. The phase transition takes place at a larger $U_c$, see Fig.~\ref{fig:sketch_phasediag}, since a repulsive next-nearest neighbor interaction makes the CDW-I unfavorable. The FRG result for $U^\prime = 0.5$ for the order parameter for different system sizes is shown in Fig.~\ref{fig:cdwi_Up0.5}. We can read off $U_c(N=42, L=10)=2.7$ and for the thermodynamic limit $U_c = 2.2$. Again, we note that for small chains, we have a larger $U_c$ and a faster increase of the order parameter. For the long chain, we observe that increasing the feedback length from $10$ to $20$ changes the result for $\langle O_\text{cdwi} \rangle$ more than it did in the $U^\prime=0$ case, cf. Fig.~\ref{fig:cdwi_cmpNL}. Thus, the convergence in $L$ has to be tested for each parameter set. However, although $L=20$ gives a slightly changed result, the read-off critical interaction is the same and we conclude that $L=10$ is also sufficient in this case.

Let us finally consider $U^\prime =1$. The small system with $N=42, \ L=10$ shows the same behavior as above, with a critical interaction of $3.7$. However, when going to a larger system size with $N=258$, we were not able to integrate the flow equations to $\Lambda_\text{f}=0$ in the ordered phase anymore. We believe that this is due to numerical difficulties. The interactions we consider, $U^\prime=1$ and $U \sim 3$, are at the verge of the strong coupling regime. Besides, the comparatively large $U^\prime$ restrains the formation of a CDW-I. But although we were not able to produce converged results in the ordered phase, the FRG can still show the onset of a different phase in the thermodynamic limit.

\subsection{Transition to CDW-II}
For dominant next-nearest neighbor interaction, a charge density wave with pattern $..110011..$ becomes favorable. In this subsection, we examine systems with a fixed $U$, i.e., a vertical line in Fig.~\ref{fig:sketch_phasediag}. For PBC, we cannot obtain a perfect CDW-II due to the restriction on the system size as discussed in Sec.~\ref{subsec:num_impl}. $N$ is not allowed to be divisible by four, and thus we either have to impose a node somewhere in the system or four occupied (unoccupied) sites next to each other. Both are unfavorable for large $U^\prime$, and thus the flow equations can not be integrated down to $\Lambda_\text{f}=0$ for arbitrary $U^\prime$. Therefore we will focus on OBC in this subsection. Then, $N$ is allowed to be a multiple of four and we get results with $\Lambda_\text{f}=0$ for all considered $U^\prime$ if we impose the correct initial symmetry breaking term as described in the last subsection. However, we need a larger trigger of $S=0.05$ to drive the system into the ordered phase. Here, we will concentrate on a small chain with $N=40$. As we have seen in the study of the phase transition from the LL to the CDW-I, where we could compare to an exact solution, the truncated FRG can only give us qualitatively correct results for the value of the critical interaction. Therefore, we do not study the dependence on $N$ here. We have also seen that the dynamic calculation yielded the same results as the static one, and we thus consider only the latter in this subsection. 
\begin{figure}
	\includegraphics[width=\linewidth]{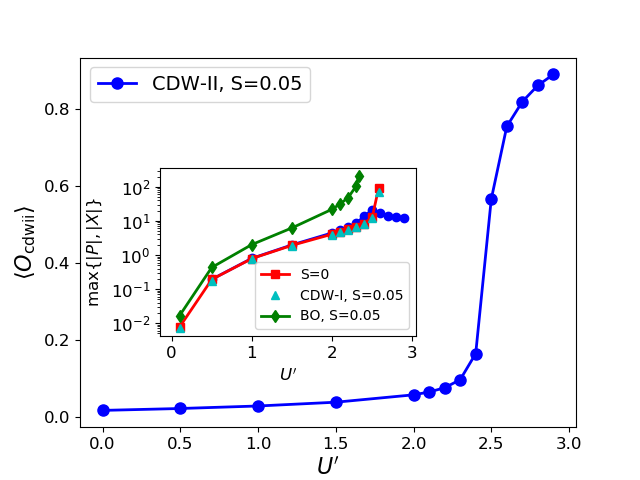}
	\caption{Phase transition to the CDW-II, see text for further explanations. OBC, $U=0, \ N=40, \ L=10$.}
	\label{fig:cdwii_OBC}
\end{figure}
In Fig.~\ref{fig:cdwii_OBC}, we show the outcome of the FRG calculation for $U=0$. If we impose an initial condition with $\Sigma_{jj}^{\Lambda_\text{i}} = (-1)^{\lfloor j /2 \rfloor} S$, we get an increase in the order parameter
\begin{equation}
	\langle O_{\text{cdwii}} \rangle = -\frac{2}{N} \sum_{j=0}^{N-1} (-1)^{\lfloor j /2 \rfloor} \langle n_j \rangle
\end{equation}
 at $U^\prime_c = 2.5$. Since $S$ is larger than in the last subsection, the interaction range over which the order parameter increases is larger.  As for the transition to the CDW-I we can ask again whether our small initial nudge to the correctly ordered system makes our ansatz inherently biased. In the inset of Fig.~\ref{fig:cdwii_OBC}, we show the maximum of the absolute value of all components of $P$ and $X$ similar to Fig.~\ref{fig:maxPX}. The different curves correspond to different initial conditions: with the correct initial condition, without a small initial symmetry breaking term, labeled as $S=0$, with an initial symmetry breaking term which nudges the system to the CDW-I phase with trigger $S=0.05$, and with an initial symmetry breaking term which biases the system to the BO phase, also with $S=0.05$.  If the system is initially nudged to the CDW-II, the maximum of the two-particle vertex stays finite for all interactions (curve labeled with ``CDW-II, $S=0.05$''). In contrast, the maximum becomes very large close to $U^\prime=2.5$ for the other initial conditions, and for interactions larger than shown we cannot integrate the flow equations down to $\Lambda_\text{f}=0$ anymore. The results for the system with initial bond order trigger are different from the other ones, and we could only obtain results up to $U^\prime=2.3$. We conclude that also in this case, the FRG correctly captures the phase transition to the CDW-II phase. The critical interaction is not as consistently determined as for the LL -- CDW-I transition, since analyzing the system with a BO bias would give a slightly changed result, but we can still give an estimate of $U_c^\prime$.

\subsection{Bond order phase}

Let us now turn to the BO phase, which is hardest to interpret and detect. In this phase, the renormalized hybridization oscillates from bond to bond. The  initial condition with the appropriate trigger is thus given by $\Sigma_{j,j \pm 1}^{\Lambda_\text{i}} = (-1)^j S$ and the order parameter by 
\begin{equation}
	\langle O_{\text{BO}} \rangle = \frac{1}{N} \sum_{j=0}^{N-1} (-1)^{j+1} \langle c_j^\dagger c_{j+1} + c_{j+1}^\dagger c_j \rangle.
\end{equation}
Since the phase transition to a BO phase arises from a competition of the kinetic energy and the next-nearest neighbor interactions, we examine systems with fixed $U$ and increasing $U^\prime$ as in the search for the CDW-II phase.

\begin{figure}
	\includegraphics[width=\linewidth]{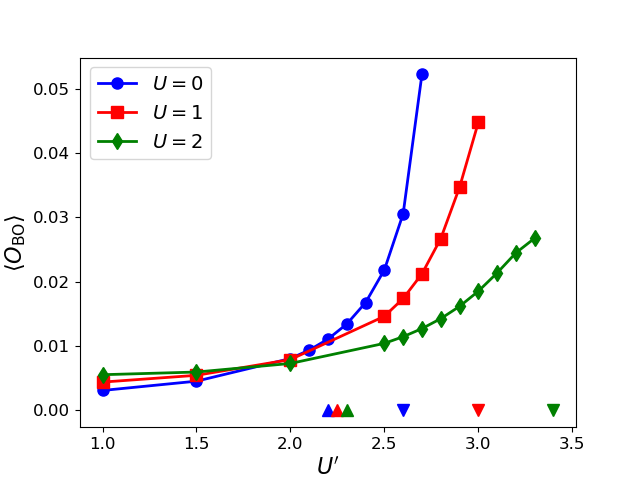}
	\caption{Phase transition from the LL to the BO phase, $N=42, \ L=10, \ S=0.001$. The triangles pointing up indicate color coded the transition LL -- BO phase, and the triangles pointing down BO -- CDW-II according to Ref.~\cite{mishra2011}.  For $U^\prime$ larger than shown in the different curves, the flow equations cannot be integrated down to $\Lambda_\text{f}=0$. }
	\label{fig:bo_PBC}
\end{figure}

We show results of the order parameter as a function of $U^\prime$ for several $U$ in Fig.~\ref{fig:bo_PBC} for a system with PBC, $N=42, \ L=10$ and initial bias $S=0.001$. We have also indicated where the phase transitions should approximately take place according to Mishra \textit{et al.} \cite{mishra2011}: Color coded for the studied nearest-neighbor interactions, the triangles pointing up indicate the transition between LL and BO, whereas the triangles pointing down show the critical $U^\prime$ at which the transition BO -- CDW-II takes place. For $U^\prime$ larger than shown, for example $U^\prime > 2.7$ for $U=0$, the flow equations can not be solved down to $\Lambda_\text{f}=0$. This reflects a phase transition to the CDW-II phase. As can be seen, the order parameter increases in the expected regions, but there is no unambiguous indication for a phase transition as was the case for the CDW phases. Besides, the flow equations can also be integrated down to $\Lambda_\text{f}=0$ in this parameter regime if no initial bias is imposed, cf. Fig.~\ref{fig:cdwii_OBC}. Thus for this phase, the FRG only gives at best inconclusive results and the BO phase would have been missed based on our FRG analysis.

We verified that a dynamic calculation does not significantly change the above presented results; also for this full second order calculation there is no clear phase transition from the LL phase to a bond order phase.

\section{Conclusion}
\label{sec:conclusion}

To conclude, we have used the truncated FRG to investigate phases in a one-dimensional fermionic model. The well-known phase transitions at intermediate interactions are especially challenging for our weak coupling approach. We used an FRG approximation scheme as in Refs.~\cite{bauer2013,weidinger2017}, which allows for an inclusion of all terms up to second order in the interaction, resulting in a dynamic self-energy and dynamic renormalized vertex functions. A feedback of the vertex on the flow is incorporated as well. This scheme can be implemented both in real space as well as in momentum space. We have shown that the resulting FRG flow equations enable us to map out charge ordered phases in the one-dimensional spinless tight-binding chain with nearest and next-nearest neighbor interactions. Since we included a flow of the self-energy, we have direct access to the order parameters. The frequency dependence can be neglected in the studied cases, static and dynamic calculations give essentially the same result. Due to our approximations, the predicted interaction strengths at which the transitions takes place are only qualitatively correct. It is also very difficult to obtain conclusions about the nature of the phase transition. An indication as clear as for the charge ordered phases is unfortunately missing for the bond order phase. We thus find that this FRG scheme can hint at ordered phases, but there is no guarantee that all of them are captured. This should be kept in mind when applying truncated FRG to other models as well. However, the detection of the CDW phases in the tight-binding model at intermediate critical interactions is a step forward compared to simple perturbation theory, mean field theory and FRG schemes that are only correct up to first order or neglect a vertex feedback. On this level of truncation and with the channel decomposition, the emerging picture is quite clear. Unfortunately, due to computational limitations, at the moment it seems very difficult to go beyond this in order to detect the bond order phase as well.

The consistent second order scheme we used offers a controlled way to obtain dynamic results as well as an inclusion of the self-energy feedback. Further applications might include the treatment of disorder \cite{karrasch2015}. We note that for the real space system with open boundary conditions, we could couple noninteracting semi-infinite leads to the interacting chain. In an FRG scheme such as above, those can be taken into account exactly without increasing the complexity of the algorithm, as was, e.g., done in Ref.~\cite{andergassen2004}. It is therefore possible to study transport properties. Besides, our real-space formulation would also allow us to study a two-dimensional system by ``folding'' the chain to form a two-dimensional lattice. The flow equations stay the same, only the range of the bare interaction would be increased. With this, we expect that we could treat systems of spinless fermions on a cubic lattice (OBC). In principle, other geometries should also be accessible.

\begin{acknowledgements}

We thank D. Rohe from the J{\"u}lich Supercomputing Centre for his support in parallelizing the code for the dynamic flow equations and C. Honerkamp for helpful discussions.
This work was supported by the Deutsche Forschungsgemeinschaft via RTG 1995 (L.M. and V.M.) and the Emmy Noether program Grant No. KA 3360/2-1 (B.S. and C.K.).
Simulations were performed with computing resources granted by RWTH Aachen University under project rwth0252. 
\end{acknowledgements}

\appendix
\setcounter{secnumdepth}{0}
\section{Appendix: Detailed derivation of the FRG flow equations}
\label{app:detail_FRG}
\renewcommand{\theequation}{A\arabic{equation}}
\setcounter{equation}{0}

We start from the general FRG flow equations, neglecting the three-particle vertex \cite{metzner2012}. For the self-energy, it holds
\begin{equation}
	\frac{ \diffd}{\diffd \Lambda} \Sigma^\Lambda (q_1^\prime, q_1) = - \frac{1}{\beta} \sum_{q_2^\prime, q_2} \mathcal{S}^\Lambda_{q_2, q_2^\prime} \, \Gamma^\Lambda(q_2^\prime, q_1^\prime; q_2,q_1).
\end{equation}
In the above equation, the indices $q_i$ are general, but in our case mean the combination of a real space index and a frequency. $\mathcal{S}^\Lambda$ denotes the single-scale propagator, which is defined as $\mathcal{S}^\Lambda = \mathcal{G}^\Lambda \left( \frac{\diffd}{\diffd \Lambda} \left[ \mathcal{G}^\Lambda_0\right]^{-1}  \right) \ \mathcal{G}^\Lambda$. This is equivalent to a derivative of the full propagator $\mathcal{G}^\Lambda$ with respect to $\Lambda$ at fixed self-energy (Sec.~\ref{subsec:FRG_gen}).

The two-point vertex can be split into different contributions, $\diffd \Gamma^\Lambda/ \diffd \Lambda = \diffd \left( \Gamma^\Lambda_p + \Gamma^\Lambda_x + \Gamma^\Lambda_d \right) /\diffd \Lambda$. Those are labeled  by ``p'' for the particle-particle channel, ``x'' for the exchange (sometimes also called crossed) particle-hole channel, and ``d'' for the direct particle-hole channel \cite{karrasch2008}. The flow equations for the three channels are given by
\begin{widetext}
\begin{align}
	\frac{ \diffd}{\diffd \Lambda} \Gamma_p^\Lambda(q_1^\prime, q_2^\prime; q_1, q_2) & = \frac{1}{ \beta} \sum_{q_3^\prime, q_3, q_4^\prime, q_4} \Gamma^\Lambda(q_1^\prime, q_2^\prime; q_3,q_4) \mathcal{S}^\Lambda_{q_3, q_3^\prime} \mathcal{G}^\Lambda_{q_4, q_4^\prime} \Gamma^\Lambda(q_3^\prime, q_4^\prime; q_1,q_2) \label{eq:flowGammap_general} \\
	\frac{ \diffd}{\diffd \Lambda} \Gamma_x^\Lambda(q_1^\prime, q_2^\prime; q_1, q_2) & = \frac{1}{ \beta} \sum_{q_3^\prime, q_3, q_4^\prime, q_4} \Gamma^\Lambda(q_1^\prime, q_4^\prime; q_3,q_2) \left[ \mathcal{S}^\Lambda_{q_3, q_3^\prime}  \mathcal{G}^\Lambda_{q_4, q_4^\prime}+  \mathcal{S}^\Lambda_{q_4, q_4^\prime}  \mathcal{G}^\Lambda_{q_3, q_3^\prime}\right] \Gamma^\Lambda(q_3^\prime, q_2^\prime; q_1,q_4)  \label{eq:flowGammax_general}  \\
	\frac{ \diffd}{\diffd \Lambda} \Gamma_d^\Lambda(q_1^\prime, q_2^\prime; q_1, q_2) & = - \frac{1}{ \beta} \sum_{q_3^\prime, q_3, q_4^\prime, q_4} \Gamma^\Lambda(q_1^\prime, q_3^\prime; q_1,q_4) \left[ \mathcal{S}^\Lambda_{q_3, q_3^\prime}  \mathcal{G}^\Lambda_{q_4, q_4^\prime}+  \mathcal{S}^\Lambda_{q_4, q_4^\prime}  \mathcal{G}^\Lambda_{q_3, q_3^\prime}\right] \Gamma^\Lambda(q_4^\prime, q_2^\prime; q_3,q_2)  \label{eq:flowGammad_general} 
\end{align}
\end{widetext}
Due to frequency conservation, we can work with only three frequencies instead of the four $(\omega_1^\prime, \omega_2^\prime; \omega_1, \omega_2)$. A natural choice are the bosonic frequencies $(\Pi, \Chi, \Delta)$ \cite{karrasch2008}
\begin{align*}
	\Pi &= \omega_1^\prime + \omega_2^\prime = \omega_1 + \omega_2 \\
	\Chi &= \omega_2- \omega_1^\prime  = \omega_2^\prime - \omega_1 \\
	\Delta & = \omega_1^\prime - \omega_1 = \omega_2 - \omega_2^\prime
\end{align*}
As stated in Sec.~\ref{subsec:FRG_floweqs}, the main idea used in Refs.~\cite{bauer2014,weidinger2017} is to insert the bare vertex $V$ in the right-hand side of the flow equations for $\Gamma_i, \ i=p,x,d$. I.~e., we replace $\Gamma^\Lambda = V + \Gamma^\Lambda_p + \Gamma^\Lambda_x + \Gamma^\Lambda_d$ by $V$. For general finite-ranged interaction, $V$ has the structure
\begin{equation}
	V(j_1^\prime, j_2^\prime; j_1, j_2) =  \delta_{j_1,j_2}^{L_U} U_{j_1, j_2} \left( \delta_{j_1, j_1^\prime} \delta_{j_2 ,j_2^\prime} - \delta_{j_1^\prime, j_2} \delta_{j_2^\prime, j_1}  \right),
\end{equation}
where $\delta_{j_1 j_2}^{L_U} = 1$ for $|j_1-j_2| \leq L_U$ with $L_U$ the range of the interaction. For the Hamiltonian given in Eq.~\eqref{eq:Hamiltonian}, it is
\begin{equation}
	U_{j_1,j_2} = U \left( \delta_{j_1,j_2+1} + \delta_{j_1,j_2-1} \right) + U^\prime \left( \delta_{j_1,j_2+2} + \delta_{j_1,j_2-2} \right)
\end{equation}
and $L_U = 2$ (or $L_U=1$ for $U^\prime=0$).

In the following, we will give a detailed derivation for the flow equation of the particle-particle channel, and only mention some intermediate results for the other channels. Inserting $V$ in the flow equation for the particle-particle channel, we get
\begin{widetext}
\begin{align}
\label{eq:flowGammap_bare}
	& \frac{ \diffd}{\diffd \Lambda} \Gamma_p^\Lambda(j_1^\prime, j_2^\prime; j_1, j_2; \Pi, \Chi, \Delta) \notag \\
	=& \delta_{j_1^\prime,j_2^\prime}^{L_U} \delta_{j_1, j_2}^{L_U} \, U_{j_1^\prime,j_2^\prime} U_{j_1, j_2} \, \frac{1}{\beta} \sum_{ \omega_3} \left[ \mathcal{S}^\Lambda_{j_1^\prime, j_1} (\omega_3) \mathcal{G}^\Lambda_{j_2^\prime, j_2} (\Pi -\omega_3) - \mathcal{S}^\Lambda_{j_1^\prime, j_2} (\omega_3) \mathcal{G}^\Lambda_{j_2^\prime, j_1} (\Pi -\omega_3) \right. \notag \\
	& \hspace{6cm} \left. - \mathcal{S}^\Lambda_{j_2^\prime, j_1} (\omega_3) \mathcal{G}^\Lambda_{j_1^\prime, j_2} (\Pi -\omega_3) + \mathcal{S}^\Lambda_{j_2^\prime, j_2} (\omega_3) \mathcal{G}^\Lambda_{j_1^\prime, j_1} (\Pi -\omega_3)\right].
\end{align}
\end{widetext}
Thus $\Gamma^\Lambda_p$ depends only on the frequency $\Pi$, and it must hold $|j_1^\prime - j_2^\prime | \leq L_U$ and $|j_1 - j_2 | \leq L_U$. This leads us to define
\begin{equation}
	P_{i,j}^{k,l; \Lambda} (\Pi) := \Gamma_p^\Lambda(i,i+k; j, j+l; \Pi). \tag{\ref{eq:defP}}
\end{equation}
From Eq.~\eqref{eq:flowGammap_bare}, we see that $i,j \in [0,N)$, while the upper indices $k,l \in [-L_U,L_U]$. For OBC, the range of the subscript indices is restricted by $\max(0,-k) \leq i < \min(N, N+k)$ and similar for $j$ with $l$ to ensure $0\leq i+k, j+l < N$. For PBC, we identify as explained in Sec.~\ref{sec:model} $-(n+1) =N-(n+1)$ or $N+n=n$ for $0 \leq n < N$, and the subscript indices lie in $[0,N)$ independent of the superscript indices. Note that for the particle-particle channel, the components with $k=0$ or $l=0$ are zero due to the antisymmetry of the vertex.

A similar consideration leads to
\begin{align}
	X_{i,j}^{k,l;\Lambda} (\Chi) & := \Gamma_x^\Lambda(i, j+l; j, i+k; \Chi)  \tag{\ref{eq:defX}}\\
	D_{i,j}^{k,l;\Lambda} (\Delta) & := \Gamma_d^\Lambda(i, j+l; i+k, j; \Delta).  \tag{\ref{eq:defD}}
\end{align}
Due to the antisymmetry of the vertex, $D=-X$ and thus we consider only the exchange particle-hole channel in the following.

So far, no vertex feedback has been included. Guided by our above calculations, we can - instead of replacing the full vertex on the right-hand side of the flow equations by only the bare interaction $V$ - replace the full vertex by a restricted one depending only on one frequency. In the flow equation for the particle-particle channel, we replace [cf. Eq.~\eqref{eq:defGammaptilde}]
\begin{align*}
	\Gamma^\Lambda (j_1^\prime \omega_1^\prime, j_2^\prime \omega_2^\prime; j_1 \omega_1, j_2 \omega_2) \rightarrow  \tilde{\Gamma}^\Lambda_p(j_1^\prime, j_2^\prime; j_1, j_2; \Pi) \notag \\
	  :=  \delta_{j_1^\prime, j_2^\prime}^L \delta_{j_1,j_2}^L \Gamma^\Lambda(j_1^\prime, j_2^\prime; j_1, j_2; \Pi,0,0).
\end{align*}
In words, $\tilde{\Gamma}^\Lambda_p$ contains the bare interaction, the full dynamic feedback of the particle-particle channel, and a static feedback from the other channels. This avoids a mixing of the bosonic frequencies \cite{jakobs2010} and ensures that the structure of the restricted channels remains valid. With this vertex feedback included, in $P_{i,j}^{k,l;\Lambda}$ also terms with $|k|,|l| > L_U$ can be generated and we restrict them to the range $[-L,L]$ with the \textit{feedback length} $L$. If we wanted to take into account the full real-space dependence of the vertex, we would have to include all terms with $k,l \in (-N/2, N/2]$ for PBC and with $k,l \in (-N,N)$ for OBC. However, as long-ranged interactions are of a higher order in the interaction, those are expected to be small and we hope that we can choose a much smaller $L$.

For a more compact notation, we introduce the matrix $\tilde{P}_{i,j}^{k,l;\Lambda}$, which we get from $\tilde{\Gamma}^\Lambda_p$ as in Eq.~\eqref{eq:defP}. An analogous consideration as above for the crossed particle-hole channel leads to a definition of $\tilde{\Gamma}^\Lambda_x$ and from this as in Eq.~\eqref{eq:defX} to  $\tilde{X}_{i,j}^{k,l;\Lambda}$.  For our model Hamiltonian Eq.~\eqref{eq:Hamiltonian}, we find
\begin{align}
	  \tilde{P}_{i,j}^{k,l; \Lambda } & (\omega) = \delta_{i,j} \delta_{k,l} \delta_{l,0}^{L_U} U_{j,j+l} - \delta_{j,i+k} \delta_{k,-l} \delta_{l,0}^{L_U} U_{j,j+l} \notag \\
	&  + P_{i,j}^{k,l; \Lambda}(\omega) + \delta_{i+k,j}^{L} \delta_{j+l,i}^L \, X_{i,j}^{(j+l-i),(i+k-j);\Lambda} (0)\notag \\
	 &  - \delta_{i,j}^L \delta_{i+k,j+l}^L \, X_{i,j+l}^{(j-i), (i+k-j-l); \Lambda} (0) \label{eq:Ptilde} \\[0.5em]
	  \tilde{X}_{i,j}^{k,l;\Lambda}& (\omega) =  \delta_{i,j} \delta_{k,l}  \delta_{k,0}^{L_U} U_{i,i+k}-\delta_{k,0} \delta_{l,0} \delta_{i,j}^{L_U} U_{j,i}  \notag \\
	& + X_{i,j}^{k,l;\Lambda} (\omega) + \delta_{i,j+l}^L \delta_{j,i+k}^L P_{i,j}^{(j+l-i),(i+k-j);\Lambda}(0)   \notag \\
	& - \delta_{i,j}^L \delta_{i+k,j+l}^L X_{i,i+k}^{(j-i),(j+l-i-k);\Lambda}(0) \label{eq:Xtilde}
\end{align}
Note that for PBC the vertex feedback has to be evaluated carefully. For example, for the feedback of the exchange particle-hole channel on the particle-particle channel, the first upper index written above as $j+l-i=:k_x$ is understood to fulfill $(i+k_x) \mod N = (j+l) \mod N$ and similar for the second upper index. Take, e.g., the component $\tilde{P}_{N-1,0}^{-1,1;\Lambda}(\Pi)$. A straightforward evaluation of the $X^\Lambda$ feedback from Eq.~\eqref{eq:Ptilde} would yield no contribution, since $j+l-i = -N+2 < -L$. Also for the second index $i+k-j=N-2>L$ and the $\delta^L$-functions are not fulfilled at first glance. However, for PBC we can go from site $i+k=N-2$ to site $j=0$ by adding $2$ and from site $j+l=1$ to site $i=N-1$ by subtracting $2$. Thus, there is a contribution from the term $X_{N-1,0}^{2,-2;\Lambda}(0)$. In contrast, for OBC, $0 \leq j+l < N$ must be fulfilled anyways, and $k_x$ can simply be computed from $j+l-i$.  

With the definitions from above, the flow equation for the particle-particle channel is now given by
\begin{align}
	& \frac{\diffd}{\diffd \Lambda} P_{i,j}^{k,l; \Lambda} (\Pi) \notag \\
	=& \frac{1}{\beta} \sum_{\cdot^\prime} \tilde{P}_{i,j^\prime}^{k,l^\prime;\Lambda} ( \Pi) \, \mathcal{S}^\Lambda_{j^\prime, i^\prime} (\omega^\prime) \mathcal{G}^\Lambda_{j^\prime+l^\prime,i^\prime+k^\prime}(\Pi -\omega^\prime) \, \tilde{P}_{i^\prime,j}^{k^\prime, l;\Lambda} ( \Pi),
\end{align}
where $\sum_{\cdot^\prime}$ indicates a summation over all primed variables, i.e., $i^\prime, j^\prime, k^\prime, l^\prime$ and also $\omega^\prime$. Defining
\begin{equation}
	W_{i,j}^{k,l;p \Lambda} (\Pi)= \frac{1}{\beta} \sum_{\omega^\prime} \mathcal{S}^\Lambda_{i, j} (\omega^\prime) \, \mathcal{G}^\Lambda_{i+k,j+l}(\Pi -\omega^\prime)
\end{equation}
we arrive at the flow equation in compact matrix multiplication form, cf. Eq.~\eqref{eq:flowP},
\begin{equation*}
	\frac{\diffd}{\diffd \Lambda} P^\Lambda (\omega)  = \tilde{P}^\Lambda(\omega) \cdot W^{p \Lambda}(\omega) \cdot \tilde{P}^\Lambda(\omega).
\end{equation*}
An analogous calculation yields the flow equation for $X$, cf. Eq.~\eqref{eq:flowX}, 
\begin{align}
\frac{\diffd}{\diffd \Lambda} X^\Lambda (\omega) & = \tilde{X}^\Lambda(\omega) \cdot W^{x \Lambda}(\omega) \cdot \tilde{X}^\Lambda(\omega), \notag \\
	W_{i,j}^{k,l;x \Lambda} (\Chi) &= \frac{1}{\beta}  \sum_{\omega}  \left[ \mathcal{S}^\Lambda_{i+k,j+l}(\omega) \, \mathcal{G}^\Lambda_{i,j}(\Chi + \omega) \right. \notag \\
	& \hspace{4em} \left. +\mathcal{S}^\Lambda_{i,j}(\Chi + \omega) \, \mathcal{G}^\Lambda_{i+k,j+l}(\omega) \right].
\end{align}
The initial condition for the vertex is $\Gamma^{\Lambda_\text{i}}=V$, such that $P^{\Lambda_\text{i}} = X^{\Lambda_\text{i}} =  0$. 

Inserting our channel-decomposed two-point vertex in the flow equation for the self-energy leads to
\begin{widetext}
\begin{align}
	\frac{\diffd }{\diffd \Lambda} \Sigma_{i,j}^\Lambda(\omega) & = - \frac{1}{\beta} \sum_{\omega^\prime} \left\{ \delta_{i,j} \sum_{k=-L_U}^{ L_U} U_{j,j+k} \mathcal{S}_{i+k,i+k}^\Lambda(\omega^\prime) - \delta_{i,j}^{L_U} U_{j,i} S_{i,j}^\Lambda(\omega^\prime) \right.+ \sum_{j_2=0}^N \sum_{l=-L}^L \mathcal{S}^\Lambda_{j_2,j_2+l}(\omega^\prime) D_{i,j_2}^{(j-i),l;\Lambda}(0) \notag \\
	& \hspace{2cm} \left. + \sum_{k=-L}^L \sum_{l=-L}^L \mathcal{S}^\Lambda_{i+k,j+l} (\omega^\prime) \left[ P_{i,j}^{k,l;\Lambda}(\omega + \omega^\prime) + X_{i,j}^{k,l;\Lambda}(\omega - \omega^\prime) \right]  \right\}
\end{align}
\end{widetext}
The initial condition $\Sigma^{\Lambda_\text{i}}$ for the Hamiltonian Eq.~\eqref{eq:Hamiltonian} when using the sharp cutoff specified below in Eq.~\eqref{eq:cutoff} requires special care \cite{enss2005}. In theory, our flow starts at $\Lambda = \infty$, but numerically we have to start at a finite value $\Lambda_\text{i} < \infty$. Due to the slow decay of the right-hand side of the flow equation for the self-energy, there is a finite contribution from the integration from $\Lambda = \infty$ to our numerical $\Lambda_\text{i}$ that we have to take into account. This problem does not occur in the flow equation of the two-particle vertex.  The above described analysis yields $\Sigma^{\Lambda_\text{i}}(\omega) = 0$ \cite{enss2005}, unless we impose a small initial symmetry breaking term.

The self-energy and the vertices fulfill the following symmetries
\begin{align}
	\Sigma_{i,j}^\Lambda (\omega) & = \Sigma_{j,i}^\Lambda(\omega) = \left[ \Sigma_{i,j} (-\omega) \right]^\star \\
	P_{i,j}^{k,l} (\Pi) &= P_{j,i}^{l,k} (\Pi) = P_{i+k,j+l}^{-k,-l} (\Pi) \notag \\
	&= -P_{i,j+l}^{k,-l} (\Pi) =\left[ P_{i,j}^{k,l} (-\Pi) \right]^\star	\\
	X_{i,j}^{k,l} (\Chi) &= X_{j,i}^{l,k} (\Chi) = X_{i+k,j+l}^{-k,-l} (-\Chi) =\left[ X_{i,j}^{k,l} (-\Chi) \right]^\star.
\end{align}
When discretizing the frequencies, we can work with $\omega \geq 0$ and infer the quantities at negative frequencies with complex conjugation.

So far, no cutoff has been imposed in the flow equations. We choose a sharp cutoff function [cf. Eq.~\eqref{eq:cutoff}] $\mathcal{G}^\Lambda_0 (\omega) = \Theta(|\omega|-\Lambda) \mathcal{G}_0 (\omega)$ such that the frequency integrals on the right-hand side can be evaluated analytically. Products of $\delta$ distributions and step functions can be computed according to Ref.~\cite{morris1994} as
\begin{equation}
\label{eq:prod_delta_theta}
	\delta_\varepsilon (x-\Lambda) f[ \Theta_\varepsilon (x-\Lambda) ] \rightarrow \delta(x-\Lambda) \int_{0}^{1} f(t) \diffd t.
\end{equation}
The expression that finally appears in the flow equations is the full Green's function where the cutoff function is no longer present, but it is still scale dependent since the self-energy depends on $\Lambda$:
\begin{equation}
	\tilde{\mathcal{G}}^\Lambda(\omega) =\left[ \left[ \mathcal{G}_0(\omega) \right]^{-1} - \Sigma^\Lambda(\omega) \right]^{-1}.
	\label{eq:propagator}
\end{equation}

For the self-energy, the only cutoff dependence comes from the single-scale propagator
\begin{equation}
	\mathcal{S}^\Lambda(\omega) = -\delta(|\omega|-\Lambda) \mathcal{G}_0(\omega) \left[1- \Sigma^\Lambda(\omega) \mathcal{G}_0(\omega) \Theta(|\omega|-\Lambda) \right]^{-2}.
\end{equation}
We thus find the following flow equation for the self-energy $\Sigma^\Lambda (\omega)$,
\begin{widetext}
\begin{align}
\label{eq:flowS}
	 \pi \frac{\diffd }{\diffd \Lambda} \Sigma_{i,i+m}^\Lambda (\omega) = &  -  \delta_{m,0} \left[ U \left( \Re \left\{ \tilde{\mathcal{G}}_{i-1,i-1}^\Lambda (\Lambda) \right\} +   \Re \left\{\tilde{\mathcal{G}}_{i+1,i+1}^\Lambda (\Lambda)\right\} \right) + U' \left (  \Re \left\{ \tilde{\mathcal{G}}_{i-2,i-2}^\Lambda (\Lambda)\right\} +   \Re \left\{ \tilde{\mathcal{G}}_{i+2,i+2}^\Lambda (\Lambda) \right\} \right) \right] \notag\\
	&  + \delta_{m,1} \, U  \,   \Re \left\{ \tilde{\mathcal{G}}_{i,i+1}^\Lambda(\Lambda) \right\} + \delta_{m,2} \, U' \,  \Re \left\{ \tilde{\mathcal{G}}_{i,i+2}^\Lambda (\Lambda) \right\} + \delta_{m,0}^L \sum_{j_2=0}^N \sum_{l=-L}^{L}  \Re \left\{ \tilde{\mathcal{G}}_{j_2,j_2+l}^\Lambda (\Lambda) \right\} X_{i,j_2}^{m,l;\Lambda} (0) \notag \\
	&  - \frac{1}{2} \sum_{\omega^\prime = \pm \Lambda} \sum_{k=-L}^{L} \sum_{l=-L}^{L} \tilde{\mathcal{G}}_{i+k,i+m+l}^\Lambda (\omega^\prime)  \left[ P_{i,i+m}^{k,l;\Lambda} (\omega+\omega^\prime) + X_{i,i+m}^{k,l;\Lambda} (\omega - \omega^\prime) \right] 
\end{align}
\end{widetext}
For $\omega=0$, the last line simplifies to
\begin{align}
\label{eq:flowS0}
	- \sum_{k,l=-L}^{L} \hspace{-0.5em} \Re \left\{ \tilde{\mathcal{G}}_{i+k,i+m+l}^\Lambda (\Lambda) \left[ P_{i,i+m}^{k,l;\Lambda} (\Lambda) + X_{i,i+m}^{k,l;\Lambda} (\Lambda)^\star \right] \right\}.
\end{align}

We can calculate the frequency integrals in the matrices $W^{p/x \Lambda}$ analytically as well. In those expressions, there are also step functions from the full Green's function next to the step functions and delta distribution of the single-scale propagator. We finally arrive at
\begin{widetext}
\begin{align}
	& W_{i,j}^{k,l;p \Lambda} (\omega)   \stackrel{\omega>0}{=} \frac{1}{2 \pi} \left\{ \left[\tilde{\mathcal{G}}^\Lambda_{i,j}(\Lambda)\right]^\star \, \tilde{\mathcal{G}}^\Lambda_{i+k,j+l}(\omega+\Lambda) + \Theta(\omega -2 \Lambda) \, \tilde{\mathcal{G}}^\Lambda_{i,j}(\Lambda) \tilde{\mathcal{G}}^\Lambda_{i+k,j+l}(\omega-\Lambda)  \right\} \label{eq:def_Wp} \\[0.5em]
	& W_{i,j}^{k,l;x \Lambda} (\omega) 	 \stackrel{\omega>0}{=} \frac{1}{2 \pi}  \left\{  \tilde{\mathcal{G}}^\Lambda_{i+k,j+l}(\Lambda) \tilde{\mathcal{G}}^\Lambda_{i,j}(\omega+\Lambda) + \left[\tilde{\mathcal{G}}^\Lambda_{i,j}(\Lambda)\right]^\star \,  \left[\tilde{\mathcal{G}}^\Lambda_{i+k,j+l}(\omega+\Lambda)\right]^\star  \right.\notag \\
	& \left. \hspace{10em} + \Theta( \omega -2 \Lambda) \left( \tilde{\mathcal{G}}^\Lambda_{i,j}(\Lambda)\, \left[ \tilde{\mathcal{G}}^\Lambda_{i+k,j+l}(\omega-\Lambda)\right]^\star + \left[\tilde{\mathcal{G}}^\Lambda_{i+k,j+l}(\Lambda)\right]^\star \,  \tilde{\mathcal{G}}^\Lambda_{i,j}(\omega-\Lambda) \right) \right\}. \label{eq:def_Wx}
\end{align}
\end{widetext}

\phantom{blub\\ \\ blub \\ }

\phantom{blub\\ \\ blub \\ }

 At zero frequency, the $W$ matrices are given by 
\begin{align}
	& W_{i,j}^{k,l;p \Lambda} (0) = \frac{1}{2 \pi} \Re \left\{ \tilde{\mathcal{G}}^\Lambda_{i,j}(\Lambda) \left[ \tilde{\mathcal{G}}^\Lambda_{i+k,j+l} (\Lambda) \right]^\star  \right\} \label{eq:def_Wp0}\\
	& W_{i,j}^{k,l;x \Lambda} (0) = \frac{1}{  \pi} \Re \left\{ \tilde{\mathcal{G}}^\Lambda_{i,j}(\Lambda)  \tilde{\mathcal{G}}^\Lambda_{i+k,j+l} (\Lambda) \right\}. \label{eq:def_Wx0}
\end{align}
We used that for $\omega >0$ (and $\Lambda \geq 0$) $\Theta(|\omega -\Lambda| -\Lambda) = \Theta(\omega - 2\Lambda)$. Note that in the calculation of $W^{p/x \Lambda}(0)$ the step functions from the Green's function have the same argument as the delta distribution and the step functions from $\mathcal{S}^\Lambda$ and have thus to be taken into account when using Eq.~\eqref{eq:prod_delta_theta}. Setting $\omega=0$ in Eqs.~\eqref{eq:def_Wp} and \eqref{eq:def_Wx} does not give the correct result.

As described in the main text in Sec.~\ref{subsec:FRG_floweqs}, we can simplify the flow equations by going to a static approximation, which then no longer contains all second order terms. To do so, we only keep the frequency $\omega=0$. Then all flowing quantities are real and only the simpler versions of the flow equations as in Eqs.~\eqref{eq:flowS0}, \eqref{eq:def_Wp0}, and \eqref{eq:def_Wx0} remain.

\bibliography{phasediagram_bib}

\end{document}